\documentclass{appolb}
\usepackage{amssymb}
\usepackage{graphicx}

\begin{document}
\pagestyle{plain}
\newcount\eLiNe\eLiNe=\inputlineno\advance\eLiNe by -1
\title{ON THE PROPAGATION OF NON STATIONARY\\
PRESSURE WAVES IN STELLAR INTERIORS%
}
\author{Patryk Mach%
\thanks{e-mail: {\tt mach@th.if.uj.edu.pl}}%
\address{Institute of Physics, Jagiellonian University,\\
Reymonta 4, 30-059~Krak\'ow, Poland}}
\maketitle

\begin{abstract}
An analysis of the propagation of non stationary waves in the adiabatic region of stellar interior is presented. An equation of motion with an effective potential is derived, similar to the Zerilli equation known in the propagation of gravitational waves. The Huyghens principle is violated in this case and the energy diffusion outward null cones is expected. Numerical calculations demonstrate that the  diffusion is weak for the case of standard Solar model; thus no significant effect corresponding to quasinormal modes can be expected. The likely reason for the absence of stronger features  is the restriction of our analysis to adiabatic regions only, where the breakdown of the Huyghens principle is insignificant.
\end{abstract}

\section{Motivation}

In the standard helioseismology (or asteroseismology, to be more general) one usually deals with stationary perturbations of the gas pressure (or density) in stellar interiors. This issue has been well investigated so far, both from the theoretical and observational side (see \eg \cite{Christensen}). Here, by stationary approach we understand posing a hydrodynamical boundary problem leading to some characteristic frequencies that can be subsequently compared with the frequencies of observed stellar oscillations. There is no reason, however, why not to consider the propagation of non stationary waves in stellar interiors. It is known that waves propagating in an inhomogeneous medium can produce some non stationary effects such as, for instance, appearance of the quasinormal modes (for an example taken from the theory of perturbations of the Schwarzschild space-time see \cite{Vishveshwara}). These are especially important from the observational point of view as their frequencies and damping coefficients are independent of the wave profile but depend on the characteristics of the medium. It is also known that in some cases the quasinormal modes can dominate \cite{Karkowski}.

In this paper we perform a simplified analysis of the problem which appears in the astrophysics of stellar interiors. The order of this work is as follows. In Section 2 we recall some basic formalism commonly used in the theory of stellar oscillations. Section 3 gives the description of the propagation of non stationary waves in the adiabatic region of the stellar interior together with the derivation of the exact form of the equation of motion. We transform this equation to the form of the Zerilli equation \cite{Zerillia}, \cite{Zerillib} which one encounters in the theory of gravitational waves propagating in the Schwarzschild background metric (a case known to give significant quasinormal modes). The Lagrangian formulation of the problem is given in Section 4 while in Section 5 we deal with the Noether's energy density and its diffusion through the characteristics. In Section 6 we present the effective potential occurring in the equation of motion, obtained for the case of standard solar model. Finally Section 7 shows the results of some numerical calculation of the propagation of non stationary pressure waves in the Sun. Some final remarks and conclusions are given in Section 8.

\section{The formalism}

In this section we shall remind some basic equations known from the Newtonian theory of stellar oscillations (or asteroseismology). We will not give any precise derivation here as it can be easily found in other papers. In turn, we will try to focus our attention mainly on putting down all the assumptions leading to the mentioned equations and we will explain all the notation we use.

We will consider the motion of gas in the star ruled by the Euler equation
\begin{equation}
\label{euler}
\varrho \partial_t \mathbf v + \varrho \mathbf v \nabla \mathbf v = - \nabla p + \varrho \mathbf g
\end{equation}
together with the continuity equation
\begin{equation}
\label{r_ciaglosci}
\partial_t \varrho + \nabla(\mathbf v \varrho) = 0.
\end{equation}
Here $\varrho$ denotes the density, $p$ the pressure and $\mathbf v$ the velocity field of the gas. The term $\mathbf g$ in the equation (\ref{euler}) stands for the gravitational acceleration. In further considerations it will be, however, much more convenient to use the gravitational potential $\Phi$ instead of $\mathbf g$. We shall assume that a non minus convention $\mathbf g = \nabla \Phi$ holds. Then, the potential $\Phi$ satisfies Poisson's equation of the form
\begin{equation}
\label{poisson}
\nabla^2 \Phi = - 4 \pi G \varrho.
\end{equation}
The above set of equations should be completed with one more, namely energy conservation equation (or the first law of thermodynamics)
\begin{equation}
\label{r_energii}
\frac{dq}{dt} = \frac{1}{\varrho (\Gamma_3 - 1)} \left( \frac{dp}{dt} -\frac{\Gamma_1 p}{\varrho} \frac{d \varrho}{dt}  \right).
\end{equation}
Here $q$ denotes specific heat (\ie heat per unit mass) and we have used standard thermodynamic notation (see \eg \cite{Cox})
\[ \Gamma_1 = \left( \frac{\partial \ln p}{\partial \ln \varrho} \right)_S, \: \Gamma_3 - 1 = \left( \frac{\partial \ln T}{\partial \ln \varrho} \right)_S. \]

The next step is to derive a set of linearized equations describing the evolution of small perturbations of the equilibrium structure of the star. Under assumptions of adiabaticity of the motion and the spherical symmetry of the undisturbed medium one may obtain: 
\begin{equation}
\label{euler_na_perturbacje_skladowa_radialna}
\varrho_0 \partial_t^2 \xi = - \partial_r p^\prime + \varrho_0 \partial_r \Phi^\prime - \varrho^\prime g_0,
\end{equation}
\begin{equation}
\label{czesc_horyzontalna_euler_wersja_ostateczna}
- \partial_t^2 \left( \varrho^\prime + \frac{1}{r^2} \partial_r (r^2 \varrho_0 \xi) \right) = - \nabla_h^2 p^\prime + \varrho_0 \nabla_h^2 \Phi^\prime,
\end{equation}
\begin{equation}
\label{r_energii_wersja_ostateczna}
\varrho^\prime = \frac{\varrho_0}{\Gamma_{1,0} p_0} p^\prime + \varrho_0 \xi \left( \frac{1}{\Gamma_{1,0}} \frac{d \ln p_0}{dr} - \frac{d \ln \varrho_0}{dr} \right),
\end{equation}
\begin{equation}
\label{poisson_wersja_ostateczna}
\frac{1}{r^2} \partial_r (r^2 \partial_r \Phi^\prime) + \nabla_h^2 \Phi^\prime = - 4 \pi G \varrho^\prime.
\end{equation}
The notation used here requires some explanation. The primed quantities denote Eulerian perturbations and are functions of both position $\mathbf r$ and time $t$, whereas quantities with zero indices describe the undisturbed medium (\ie an equilibrium structure of the star) and are only functions of the distance from the center of the star, due to the spherical symmetry we have assumed. Thus, for instance
\begin{eqnarray}
\varrho(\mathbf r, t) & = & \varrho_0(r) + \varrho^\prime(\mathbf r,t), \nonumber \\
p(\mathbf r,t) & = & p_0(r) + p^\prime(\mathbf r,t), \dots
\label{perturbacje_eulera}
\end{eqnarray}
By $\xi$ we have denoted the radial part of the gas displacement, \ie
\begin{equation}
\label{perturbacja_lagrangea_r}
\delta \mathbf r = \xi \mathbf e_r + \mathbf \xi_h,
\end{equation}
where $\mathbf e_r$ stands for the unit vector in the radial direction and $\mathbf \xi_h$ is a horizontal part of the displacement vector. Finally $\nabla_h$ denotes the horizontal part of the gradient operator. The reader interested in rigorous derivation of the above equations may consult \cite{Christensen}.

\section{Non stationary perturbations, effective potential}

The equations introduced in the preceding section possess a class of stationary solutions, which is actually one of the main interests of the theory of stellar oscillations. We may, however, try to look for the non stationary solutions that could, in fact, have some physical meaning. The aim of this section is to derive some simplified equations in the form suitable for further numerical search for such solutions.

We proceed with the separation of variables. All perturbations of our interest, such as $\xi$, $p^\prime$, $\varrho^\prime$ may be expanded in the series of spherical harmonics. To simplify the notation we will drop the adequate spherical harmonics indices in the amplitudes. Due to the linearity of the obtained equations it is sufficient to write
\begin{eqnarray*}
\xi(r,\theta,\phi,t) & = & \tilde \xi(r,t) Y_{lm}(\theta,\phi),\\
p^\prime(r,\theta,\phi,t) & = & \tilde p(r,t) Y_{lm}(\theta,\phi), \dots
\end{eqnarray*}
Taking into the account that
\[ \nabla_h^2 Y_l = - \frac{l(l+1)}{r^2}Y_l, \]
and substituting the above expressions to the equations (\ref{euler_na_perturbacje_skladowa_radialna}), (\ref{czesc_horyzontalna_euler_wersja_ostateczna}), (\ref{r_energii_wersja_ostateczna}) and (\ref{poisson_wersja_ostateczna}) we get
\begin{equation}
\label{euler_czesc_radialna_po_separacji}
\varrho_0 \partial_t^2 \tilde \xi = - \partial_r \tilde p + \varrho_0 \partial_r \tilde \Phi - \tilde \varrho g_0,
\end{equation}
\begin{equation}
\label{euler_czesc_horyzontalna_po_separacji}
- \partial_t^2 \tilde \varrho - \frac{1}{r^2} \partial_r (r^2 \varrho_0 \partial_t^2 \tilde \xi) = \frac{l(l+1)}{r^2} (\tilde p - \varrho_0 \tilde \Phi),
\end{equation}
\begin{equation}
\label{poisson_po_separacji}
\frac{1}{r^2} \partial_r (r^2 \partial_r \tilde \Phi) - \frac{l(l+1)}{r^2} \tilde \Phi = - 4 \pi G \tilde \varrho
\end{equation}
and
\begin{equation}
\label{r_energii_po_separacji}
\tilde \varrho = \frac{\varrho_0}{\Gamma_{1,0} p_0} \tilde p + \varrho_0 \tilde \xi \left( \frac{1}{\Gamma_{1,0}} \frac{d \ln p_0}{dr} - \frac{d \ln \varrho_0}{dr} \right).
\end{equation}

The mentioned stationary solutions can now be obtained by setting
\[ \tilde f(r,t) = \hat f(r) e^{-i \omega t} \]
for each amplitude $\tilde f(r,t)$ of a hydrodynamical variable $f$. By completing ordinary differential equations obtained this way with the suitable boundary conditions we can determine characteristic frequencies $\omega$ of the oscillation modes.

We will, however, try to proceed in a different way. Instead of posing an eigenvalue, boundary problem, we will try to formulate some Cauchy problem with just one, second order, partial differential equation, describing the time evolution of an initial perturbation. Additionally, we will not consider any boundary conditions and thus we will treat a star as a formally infinitely distributed medium. Of course, we will not consider the propagation of the perturbations beyond the assumed radius of the star.

We shall now formulate two important, simplifying assumptions:
\begin{enumerate}
\item We will put $\tilde \Phi \equiv 0$ everywhere. This assumption was examined for the first time by Cowling. It means simply neglecting the changes in the gravitational field that arise due to small perturbations of the density in the star.
\item We will assume that the medium is adiabatic, \ie
\begin{equation}
\label{adiabatycznosc}
\frac{1}{\Gamma_{1,0}}\frac{d \ln p_0}{dr} - \frac{d \ln \varrho_0}{dr} = 0.
\end{equation}
With this assumption we eliminate the propagation of the internal gravity waves, focusing attention only on the pressure waves in the star. Now, of course, we will have to examine, whether the star region in which the perturbation propagates is indeed adiabatic. The convective zone in the stellar atmosphere can serve as an example of such region.
\end{enumerate}

Under above assumptions the considered set of equations reduces to the following form
\begin{equation}
\label{euler_skladowa_horyzontalna_po_zalozeniach}
-\partial_t^2 \tilde \varrho - \frac{1}{r^2} \partial_r(r^2 \varrho_0 \partial_t^2 \tilde \xi) = \frac{l(l+1)}{r^2}\tilde p,
\end{equation}
\begin{equation}
\label{euler_skladowa_radialna_po_zalozeniach}
\varrho_0 \partial_t^2 \tilde \xi = - \partial_r \tilde p - \tilde \varrho g_0,
\end{equation}
\begin{equation}
\label{r_energii_po_zalozeniach}
\tilde \varrho = \frac{\varrho_0}{\Gamma_{1,0} p_0} \tilde p.
\end{equation}
We may now put the equation (\ref{r_energii_po_zalozeniach}) into the equation (\ref{euler_skladowa_horyzontalna_po_zalozeniach}) to obtain
\[ - \frac{\varrho_0}{\Gamma_{1,0} p_0} \partial_t^2 \tilde p - \frac{1}{r^2} \partial_r (r^2 \varrho_0 \partial_t^2 \tilde \xi) = \frac{l(l+1)}{r^2} \tilde p. \]
The term $\partial_t^2 \tilde \xi$ in the last equation can be subsequently eliminated using (\ref{euler_skladowa_radialna_po_zalozeniach}). Taking into account that $- g_0 \varrho_0 = dp_0/dr$ and remembering the condition (\ref{adiabatycznosc}) we get after some calculations
\begin{eqnarray*}
 - \frac{\varrho_0}{\Gamma_{1,0} p_0} \partial_t^2 \tilde p + \partial_r^2 \tilde p + \left( \frac{2}{r} - \frac{d \ln \varrho_0}{dr} \right) \partial_r \tilde p + & & \\
- \left( \frac{2}{r} \frac{d \ln \varrho_0}{dr} + \frac{d^2 \ln \varrho_0}{dr^2} + \frac{l(l+1)}{r^2} \right) \tilde p & = &  0.
\end{eqnarray*}
We can now get rid of the term containing the derivative $\partial_r \tilde p$ by introducing a new dynamical variable $P$, defined with the equation
\[ \tilde p = \frac{\sqrt{\varrho_0}}{r} P. \]
Indeed, after some calculations we obtain an equation of the form
\begin{equation}
\label{r_ruchu_na_p}
 - \frac{1}{c^2(r)} \partial_t^2 P + \partial_r^2 P -V(r)P = 0,
\end{equation}
where
\begin{equation}
\label{predkosc_dzwieku}
c^2(r) = \frac{\Gamma_{1,0} p_0}{\varrho_0}.
\end{equation}
By $V$ we have denoted here a function playing the role of an effective potential
\begin{equation}
V(r) = \frac{1}{r}\frac{d \ln \varrho_0}{dr} + \frac{1}{2}\frac{d^2 \ln \varrho_0}{dr^2} + \frac{1}{4} \left( \frac{d \ln \varrho_0}{dr} \right)^2 + \frac{l(l+1)}{r^2}. \label{potencjal}
\end{equation}
Equation (\ref{r_ruchu_na_p}) can be still transformed into even more convenient form. We will eliminate the term $c^{-2}$ standing before the time derivative of $P$ by introducing a new coordinate
\begin{equation}
\label{r_gwiazdka}
r^\ast = \int\limits_{0}^{r} \frac{dr^\prime}{c(r^\prime)}.
\end{equation}
Thus an obvious relation
\[ \frac{dr^\ast}{dr} = \frac{1}{c(r)} \]
holds and the equation (\ref{r_ruchu_na_p}) may be written as
\[ - \partial_t^2 P + \partial_{r^\ast}^2 P - \frac{d \ln c}{dr^\ast} \partial_{r^\ast} P - c^2VP =
 0. \]
We will get rid of the term proportional to $\partial_{r^\ast} P$ in a way similar to that we had used before. We define a function $\Pi$ with a relation
\[P = \sqrt{c} \Pi \]
to derive the final version of our equation of motion
\begin{equation}
\label{r_ruchu_na_pi}
 -\partial_t^2 \Pi + \partial_{r^\ast}^2 \Pi - \tilde V \Pi = 0
\end{equation}
in which a new effective potential
\begin{equation}
\label{potencjal_efektywny}
\tilde V = c^2 V + \frac{1}{4}\left( \frac{d \ln c}{d r^\ast} \right)^2 - \frac{1}{2} \frac{d^2 \ln
 c}{dr^{\ast 2}}
\end{equation}
has been introduced.

\section{Lagrangian description, energy}

One may notice that the equation (\ref{r_ruchu_na_pi}) can be obtained from the variational principle
\[ \delta_\Pi S = 0 \]
by taking an action $S$ of the form
\begin{equation}
\label{dzialanie}
S = \int \mathcal L dtdr^\ast = - \frac{1}{2} \int \left( (\partial_t \Pi)^2 - (\partial_{r^\ast}\Pi)^2 - \tilde V \Pi^2 \right)dtdr^\ast .
\end{equation}
The equation (\ref{r_ruchu_na_pi}) appears then as the Euler--Lagrange equation for the Lagrangian density  $\mathcal L$, \ie
\begin{equation}
\label{euler_lagrange}
\partial_\Pi \mathcal L - \partial_t \frac{\partial \mathcal L}{\partial \partial_t \Pi} - \partial_{r^\ast} \frac{\partial \mathcal L}{\partial \partial_{r^\ast} \Pi} = 0.
\end{equation}

We can now make use of the first Noether theorem applied to the action (\ref{dzialanie}), what should allow us to define an energy for the $\Pi$ amplitudes. We will present this issue in a little detail due to some subtle matters that appear here.

We begin with considering an infinitesimal time translation of some domain $\Omega$
\[ \psi \colon \mathbb R^2 \supset \Omega \ni (t,r) \mapsto (t^\prime,r) = (t - \varepsilon,r) \in \mathbb R^2, \]
which will be assumed to be a symmetry which means that a variation of the action (\ref{dzialanie}) caused by this transformation vanishes in the domain $\Omega$. If in addition we assume that the motion happens to be real (the Euler--Lagrange equation (\ref{euler_lagrange}) is satisfied) then after some calculations we obtain
\begin{equation}
\delta S = \varepsilon \int_{\partial \Omega} \left( \left( \mathcal L - \frac{\partial \mathcal L}{\partial \partial_t \Pi} \partial_t \Pi \right) dr^\ast - \left( \frac{\partial \mathcal L}{\partial \partial_{r^\ast} \Pi} \partial_t \Pi \right)dt \right) = 0.
\label{efekty_brzegowe}
\end{equation}
The amplitude $\Pi$ is defined in a half plane $r^\ast \geqslant 0$. As a domain $\Omega$, over which we proceed with integration we may now take a part of that half plane enclosed between two constant time lines, given by the equations  $t = t_1$ and $t = t_2$.

Let us notice, that the amplitudes $\Pi$ have appeared in the separation of variables in the $(3+1)$ dimensional problem and, therefore, have to satisfy some additional conditions. In particular $\tilde p$ needs to be finite in its domain and thus also at $r^\ast = r = 0$. It follows simply that an equality $\Pi(r^\ast = 0, t) = 0$, and so  $\partial_t \Pi(r^\ast = 0, t) = 0$ must hold. Therefore, for the $\Omega$ chosen above we have
\[ \int_{\partial \Omega} \left( \frac{\partial \mathcal L}{\partial \partial_{r^\ast} \Pi} \right)dt = \int\limits_{t_1}^{t_2} \left( \frac{\partial \mathcal L}{\partial \partial_{r^\ast} \Pi} \partial_t \Pi \right)_{r^\ast = 0} dt = 0. \]
Finally it follows that the quantity
\[ \int\limits_0^\infty \left( \mathcal L - \frac{\partial \mathcal L}{\partial \partial_t \Pi} \partial_t \Pi \right)dr^\ast \]
is conserved, \ie constant in time. The expression
\begin{equation}
\mathcal E = \mathcal L - \frac{\partial \mathcal L}{\partial \partial_t \Pi} \partial_t \Pi = \frac{1}{2} \left( (\partial_t \Pi)^2 + (\partial_{r^\ast} \Pi)^2 + \tilde V \Pi^2 \right)
\label{gestosc_energii}
\end{equation}
can thus be interpreted as an energy density.

In consistency with the comment made earlier, we have assumed here that a medium in which the waves propagate is infinite and all perturbations vanish at least at infinity.

\section{Energy diffusion}

\begin{figure}
\begin{center}
\includegraphics{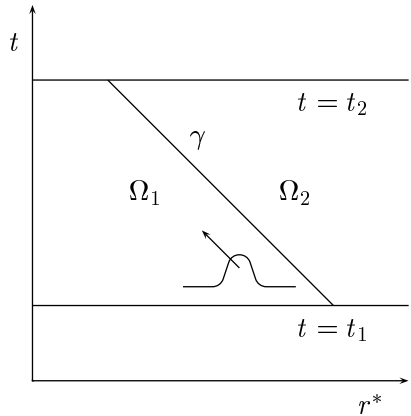}
\end{center}
\caption{An illustration of the scattering process considered in this paper.}
\label{schemat}
\end{figure}

Let us consider now $\gamma$, being a part of the incoming characteristic of the equation (\ref{r_ruchu_na_pi}), with an origin in the point with the coordinates $r^\ast = r_1$, $t = t_1$. The characteristic $\gamma$ divides the domain $\Omega$ enclosed between two constant time lines $t = t_1$ and $t = t_2$ into two subdomains: the inner one, $\Omega_1$, and the outer one $\Omega_2$ (Fig.~\ref{schemat}). Let us consider next a point with the coordinate $r^\ast = R$ lying on the $\gamma$. The expression
\begin{equation}
\label{prad_energii}
j(R,t) = (-\partial_R + \partial_t) \int\limits_{R}^\infty \mathcal E dr^\ast
\end{equation}
may be interpreted as a rate of the energy change along $\gamma$. A straightforward calculation making use of equation (\ref{gestosc_energii}) and of the motion equation (\ref{r_ruchu_na_pi}) shows that
\[ j(R,t) = \frac{1}{2} \left( (\partial_t \Pi - \partial_{r^\ast} \Pi)^2 + \tilde V \Pi^2 \right)_{r^\ast = R}. \]
Calculations leading to the above result may be simplified even more by noticing that
\[ \partial_t \frac{1}{2} \int\limits_R^\infty \mathcal E dr^\ast = \left(- \frac{\partial \mathcal L}{\partial \partial_{r^\ast} \Pi} \partial_t \Pi  \right)_{r^\ast = R}, \]
what in fact we had obtained earlier by writing formula (\ref{efekty_brzegowe}).

The whole energy $\Delta E$, that diffused from a domain $\Omega_1$ to the domain $\Omega_2$ during a time between $t_1$ and $t_2$ may be calculated as an integral of the expression (\ref{prad_energii}) over $\gamma$
\begin{eqnarray}
\Delta E & = & \int\limits_{t_1}^{t_2}  j(R,t)_{R = - t + r_1 + t_1 } dt = \label{energia_rozproszona} \\
& = & \int\limits_{t_1}^{t_2} \frac{1}{2} \left( (\partial_t \Pi - \partial_{r^\ast} \Pi)^2 + \tilde V \Pi^2 \right)_{r^\ast = - t + r_1 + t_1} dt . \nonumber
\end{eqnarray}

Let us now imagine that an incoming perturbation of compact support enclosed initially in the domain $\Omega_1$ moves along the characteristic $\gamma$. The quantity $\Delta E$ given by the expression (\ref{energia_rozproszona}) would represent the whole energy scattered outwards (into the domain $\Omega_2$) during the period between $t_1$ and $t_2$. This scattering is mathematically due to the non-vanishing potential $\tilde V$ and, to be more precise, to a potential that differs from a single term proportional to the $l(l+1)/r^2$ which will always arise from the separation of variables in the wave problem of spherical symmetry.

Experience tells us that robust energy diffusion signals the presence of quasinormal modes. Since quasinormal modes are rather difficult to be found numerically, we prefer to deal with examining the energy diffusion instead \cite{Karkowski}.

In the next sections of this paper we will concern ourselves with examining such energy scattering for a realistic case, namely standard solar model.

\section{Effective potential for a standard solar model}

We will now apply the results of the preceding sections to a standard solar model. We have already stated, that a fundamental assumption of our simplified model is that of adiabaticity of the region in which the waves propagate, and that validity of this assumption need to be carefully examined. The standard solar model has been chosen because of the existence of the convective zone in which the condition (\ref{adiabatycznosc}) is satisfied  up to a high degree.

\begin{figure}
\begin{center}
\resizebox{9cm}{!}{\includegraphics{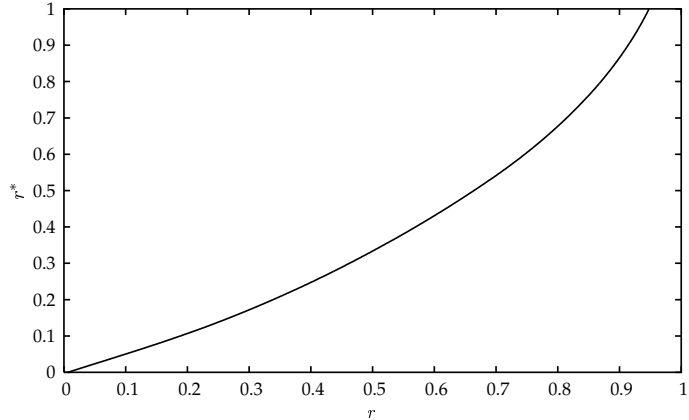}}
\end{center}
\caption{The $r^\ast$ coordinate plotted \textit{versus} radius $r$. The $r$ values are expressed in the radius of the Sun units, while $r^\ast$ is just arbitrarily normalized.}
\label{r_gwiazdka_wykres}
\end{figure}

\begin{figure}
\begin{center}
\resizebox{9cm}{!}{\includegraphics{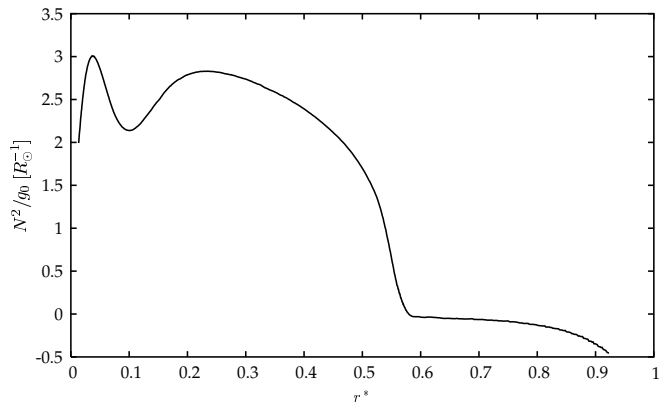}}
\end{center}
\caption{The values of $N^2/g_0$, $N$ being the Brunt--V\"{a}is\"{a}l\"{a} frequency, obtained for the used solar model and plotted \textit{versus} $r^\ast$.}
\label{adiabatycznosc_wykres}
\end{figure} 

\begin{figure}
\begin{center}
\resizebox{9cm}{!}{\includegraphics{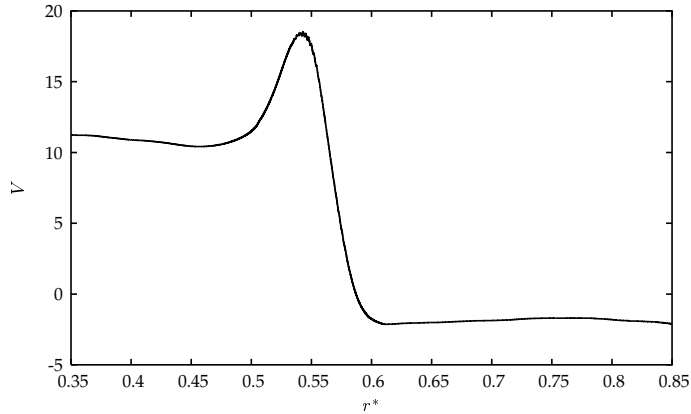}}
\end{center}
\caption{The effective potential $\tilde V_0$ for the used solar model plotted as a function of $r^\ast$.}
\label{potencjal_wykres}
\end{figure}

All our numerical calculations were made on the basis of a standard solar model computed by Bahcall, Pinsonneault \& Basin \cite{Bahcall}. The obtained relation between the variable $r^\ast$ and the radius $r$ is shown on the Fig.~\ref{r_gwiazdka_wykres}. It should be noticed here, that the $r^\ast$ variable was normalized in such a way that it changes between the values 0 and 1 for the used data range. Next, the function
\[\frac{N^2}{g_0} = \frac{1}{\Gamma_{1,0}}\frac{d \ln p_0}{dr} - \frac{d \ln \varrho_0}{dr} \]
is plotted \textit{versus} $r^\ast$ in Fig.~\ref{adiabatycznosc_wykres}. We have adopted the notation $N^2/g_0$ here, as $N$ corresponds to the well known Brunt--V\"{a}is\"{a}l\"{a} frequency. This plot shows, that we may regard the area with approximately $r^\ast \gtrapprox 0.58$ as satisfying our assumption of adiabaticity. The effective potential, i.e. $\tilde V$ for $l = 0$ is, in turn, plotted in Fig.~\ref{potencjal_wykres}. Unfortunately the only one interesting feature of this potential, that is a clear peak at $r^\ast \sim 0.5$, remains outside the adiabaticity area. Therefore we can not consider any effects caused by the existence of this peak as being physically meaningful.

One remark should be made here concerning the figures presented in this section. It is not possible to differentiate the data presented by Bahcall \textit{et al.}. in any straightforward way to obtain any smooth enough functions and therefore some smoothing procedure appears to be necessary. A slight modification of the so called Savitzky--Golay filter (see \eg \cite{nr}) was used here to obtain presented result.

\section{Example of some numerical calculations}

\begin{figure}
\begin{center}
\resizebox{9cm}{!}{\includegraphics{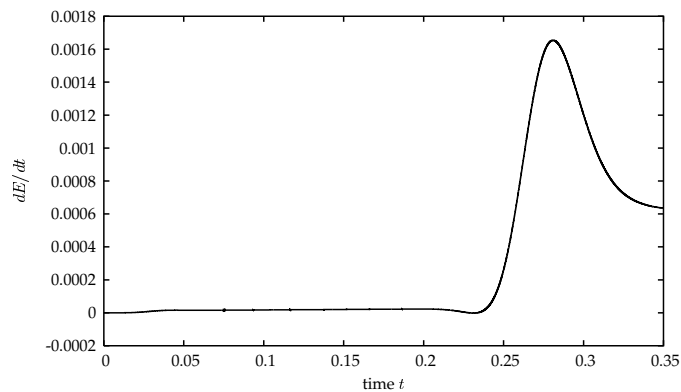}}
\end{center}
\caption{The time derivative of the energy scattered backwards. Here energy is expressed in the units of total energy of the initial perturbation, which is actually conserved, as described in section 4.}
\label{wyniki_wykres}
\end{figure}

Finally it was possible to examine, how an initially incoming wave package evolves. Fig.~\ref{wyniki_wykres} presents an example result of the numerical experiment explained in Section 5.

For simplicity, only a spherically symmetric perturbation \ie the case with $l=0$ is considered here. Propagation of all other modes may be examined in exactly the same way by taking the potential $\tilde V$ for an arbitrary $l$ number.

Here, the characteristic $\gamma$ was chosen to originate at $r^\ast = 0.845$ and the initial (\ie for $t=0$), purely incoming perturbation was taken to be a function of the shape defined by
\[ \Pi (r^\ast, t=0) = \left\{
\begin{array}{cl}
A \sin^2 \left(\frac{\pi(r^\ast - a)}{b-a}\right), &\mbox{if } r^\ast \in [a,b],\\
0, &\mbox{otherwise,}
\end{array}
\right. \]
thus being just one, bell like part of the squared sine function, centered on a compact support $[a,b]$. This is, of course, a continuous and differentiable function.

We have also examined the case with an initial perturbation in the form of standard $C^\infty$ class function of a compact support. Such function may be constructed in a well known way with an use of the exponential function. It has also a bell like shape but it gives lower FWHM to support length ratio. It appears that the squared sine function is much more useful for our purpose as, basically, the amount of scattered energy increases with an increase of the FWHM of the initial perturbation.

In the presented case, $a$ and $b$ were given the values $a = 0.73$ and $b = 0.845$ which correspond to the initial data support lying entirely in the $\Omega_1$ domain (see Fig.~\ref{schemat}). The plot presented on Fig.~\ref{wyniki_wykres} shows the time derivative of the energy that has diffused from the domain $\Omega_1$ to the outside domain $\Omega_2$ divided by the whole initial perturbation energy. The variable $r^\ast$ and the units of $t$ (approximately 2500 s per unit) are defined in such a way that the sound speed expressed in these coordinates equals unity. Thus, looking at Fig.~\ref{wyniki_wykres} it is easy to see how far could the initial perturbation arrive for a given time. Clearly, the large peak in the scattered energy corresponds to the mentioned bump in the effective potential which, as it has been already stated, cannot be considered in a convincing way as giving any results of physical meaning. It is, however, a good example of effects that may arise in the propagation of the non stationary waves due to the inhomogeneity of the medium.

\begin{figure}
\begin{center}
\resizebox{9cm}{!}{\includegraphics{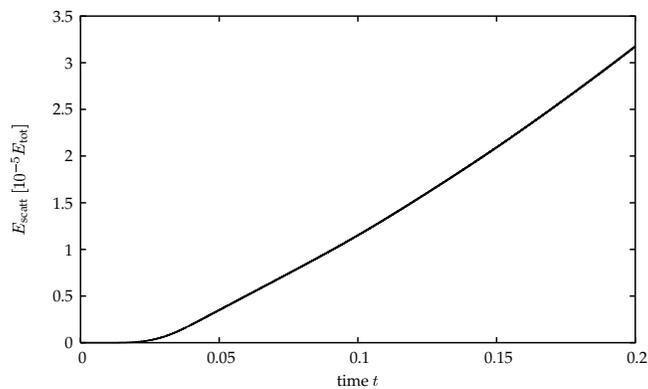}}
\end{center}
\caption{The energy $E_\mathrm{scatt}$ scattered through the characteristic expressed as a function of time $t$ for $t < 0.2$. Here $E_\mathrm{tot}$ denotes total initial energy of the wave package.}
\label{calka_wykres}
\end{figure}

In our considerations we are of course restricted to the area where the adiabaticity assumption is satisfied. In fact, even in this area some diffusion of energy does occur but on a negligible scale. Fig.~\ref{calka_wykres} shows the energy that has diffused through the characteristic as a function of time. These are in fact the same data which we have already plotted on the Fig.~\ref{wyniki_wykres} but this time restricted to the times lower than 0.2 what corresponds to the propagation in the adiabatic zone.

\section{Final remarks}

It is already well known that non stationary waves can carry interesting information about an inhomogeneous medium in which they propagate \cite{Vishveshwara}, \cite{Zerillia}, \cite{Zerillib}, \cite{Karkowski}. In this paper we have examined the propagation of such waves in the Solar convective zone by looking at the energy diffusion process.  It appears that the energy scattering occurs with rather negligible efficiency and, consequently, we expect the quasinormal modes to be absent. This may, however, follow from the fact that we have restricted ourselves to the, perhaps not interesting, simplified case of adiabatic media. It is possible that the investigation of the full model which takes into account all important physical aspects would lead to some positive results. It is also possible that positive results can be obtained by repeating the calculations presented in this paper for the models of some other stars.

\section*{Acknowledgements}

I wish to thank Professor Edward Malec for showing me the issue of non stationary waves effects and his great help in doing this work.


\begin{thebibliography}{aaa}

\bibitem{Bahcall} J. N. Bahcall, M. H. Pinsonneault, S. Basin, \textit{ApJ}. \textbf{555}, 990 (2001).
\bibitem{Christensen} J. Christensen--Dalsgaard, \textit{Lecture Notes on Stellar Oscillations}, {\tt http://astro.phys.au.dk/\~{}jcd/oscilnotes/}, 2003.
\bibitem{Cox} J. P. Cox, R. T. Giuli, \textit{Principles of Stellar Structure}, Gordon and Breach, New York 1968.
\bibitem{Karkowski} J. Karkowski, K. Roszkowski, Z. \'{S}wierszczy\'{n}ski, E. Malec, \textit{Phys. Rev.} \textbf{D 67}, 064024 (2003).
\bibitem{nr} W. H. Press, S. A. Teukolsky, W. T. Vetterling et al., \textit{Numerical Recipes in~C}, Cambridge University Press, 2002.
\bibitem{Vishveshwara} C. V. Vishveshwara, \textit{Phys. Rev.} \textbf{D 1}, 2870 (1970).
\bibitem{Zerillia} F. J. Zerilli, \textit{Phys. Rev. Lett.} \textbf{24}, 737 (1970).
\bibitem{Zerillib} F. J. Zerilli, \textit{Phys. Rev.} \textbf{D 2}, 2141 (1970).

\end{thebibliography}
\end{document}